# Phonon Spectra in the Parent Superconducting Iron-tuned Telluride $Fe_{1+x}Te$ from Inelastic Neutron Scattering and Ab Initio Calculations

*Mohamed Zbiri*[a,*] and *Romain Viennois*[b,c]

[a]Institut Laue-Langevin, 71 avenue des Martyrs, Grenoble Cedex 9, 38042, France

[b]Institut Charles Gerhardt Montpellier, UMR 5253, Université Montpellier 2 and CNRS, 34095 Montpellier, France

[c]Département de Physique de la Matière Condensée, Ecole de Physique, Université de Genève, 1211 Genève, Switzerland

## Abstract

We report inelastic neutron scattering measurements of phonon spectra in the parent superconductor iron-tuned chalcogenide $Fe_{1+x}Te$, for two different x contents ($x \leq 0.11$), using neutron time-of-flight technique. Thermal neutron spectroscopy allowed to collect the low-temperature Stokes spectra over an extended $Q$-range, at 2, 40 and 120K - hence covering both the magnetic monoclinic and the paramagnetic tetragonal phases. Whereas cold-neutrons allowed to measure high-resolution anti-Stokes spectra at 140, 220 and 300K, thus covering the tetragonal phase. Our results evidence a spin-phonon coupling behaviour towards the observed noticeable temperature-dependent change of the Stokes spectra across the transition temperatures. On the other hand, the anti-Stokes spectra reveal a pronounced hardening of the low-energy, acoustic region, of the phonon spectrum, upon heating, indicating a strong anharmonicity and a subtle dependence of phonons on structural evolution within the tetragonal phase. Experimental results are accompanied by ab initio calculations of phonon spectra of the tetragonal stoichiometric phase for a comparison with the high-resolution anti-Stokes spectra. Calculations included different density functional methods. Spin polarization and van der Waals interaction, were either considered or neglected, individually or concomitantly, in order to study their respective effect on lattice dynamics description. Our results suggest that including van der Waals interaction has only a slight effect on phonon dynamics, however, phonon spectra are better described when spin polarization is included, in a cooperative way with van der Waals interactions.

*zbiri@ill.fr

# I. Introduction

Understanding the structural and dynamical properties of high-temperature superconductors is an important step towards identifying the ultimate process of the electron pairing mechanism [1-6]. Iron-based chalcogenides offer a suitable class of materials to reach this goal, combining a simple structural framework with a relatively high-temperature superconducting transition temperature, $T_c$, achieved either chemically by a controlled doping, or mechanically by applying pressure [3,6-8]. The interest is further motivated following the recent observation of an enhanced $T_c$ as high as 50 K, and even reaching 100 K in single layer of un-doped FeSe deposited on $SrTiO_3$ substrate [3,8-10] - offering exceptionally the highest observed $T_c$ for materials other than cuprates. In this context, the parent superconductor iron-tuned telluride, $Fe_{1+x}Te$, attracted a keen interest and many studies were devoted to probe its microscopic as well as macroscopic properties [6-8,11-23]. The nuclear and magnetic structures of $Fe_{1+x}Te$ with different controlled iron excess contents were studied by different authors who determined the underlying characteristics as a function of temperature [11-14], pressure [15] and magnetic field [16,17]. Magnetically, following the amount of the interstitial iron content, x, different spin ordering configurations can be induced. Thus, for a low interstitial concentration (x < 0.11), a bi-collinear commensurate AFM phase is observed, while for a higher interstitial content (x > 0.13), a spiral incommensurate AFM order occurs. Interestingly, ferromagnetism is also observed upon applying pressure higher than 1 GPa [15], offering a unique characteristic among the iron-based superconductor family. The presence of interstitial iron atoms is a key ingredient affecting the emergence of bulk superconductivity in $Fe_{1+x}Te$ [2,6-8,18,19]. However, it is worth to notice that superconductivity is also observed up to 13 K in strained thin films of $Fe_{1+x}Te$, and whose origin is still unclear as its filamentary nature is induced by incorporating oxygen into the bulk phase of $Fe_{1+x}Te$ [20-23]. The expected coupling/interplay between structural and magnetic degrees of freedom in this material offers a magneto-elastic coupling framework as observed in other iron-based compounds [24-26]. In this context, Mossbauer measurements on Fe(Te,Se) point also towards a correlated picture gathering superconductivity, spin degrees of freedom and lattice properties [27]. Particularly, in the presently reported case of bulk $Fe_{1+x}Te$, with x ≤ 0.11, the possible coupling between the lattice and spin degrees of freedom is further supported by the occurrence of a magnetic field driven first-order simultaneous structural and magnetic transition, which was explained in terms of a detwinning due to the application of magnetic field [16,17], and the emergence of a (high) pressure induced ferromagnetic ground state [15]. Consequently, a better understanding of the magneto-elastic coupling mechanisms and properties of $Fe_{1+x}Te$ imposes to investigate the lattice dynamics side. In this context, Raman and infrared spectroscopies were used to map out the frequency shifts, and to explore the related vibrational features [28-33]. However, these techniques are subject to selection rules and are restricted only to the zone-centre region of the Brillouin zone. On the computational side, DFT calculations of the lattice dynamics of iron telluride, at the Γ-point, were also reported [28-33]. However, these were found to often disagree with each other, and with the available Γ-point experiments, when considering fully relaxed structures (both lattice parameters and atomic coordinates), so that a reasonable comparison with experimental data was performed using partially relaxed geometries (relaxing only atomic coordinate and fixing the lattice parameters at their experimentally-refined values).

This was possibly due to a strong dependence of the structural parameters on the exchange-correlation functionals, such an aspect has not systematically been investigated, concerning lattice dynamics in iron chalcogenides. Further, and also adopting a partially relaxed structure, Li and coworkers reported a computational study of both the electronic and phonon (dispersions and density of states) structures of stoichiometric FeTe (without excess Fe), and compared outcomes of their magnetic and non-magnetic calculations [34]. It is also worth mentioning an early computational work by Subedi and Singh, but dealing with lattice dynamics of the sister compound iron selenide, FeSe [35].

Here, we propose to go a step further by measuring phonon spectra in iron-tuned telluride, using inelastic neutron scattering (INS) technique, allowing overcoming the above limitations by probing, without any selection rule, the whole Brillouin zone. In this paper we have used INS to yield information on the phonon density of states in $Fe_{1+x}Te$. We present the temperature-dependence of phonon spectra measured for two different nominal iron contents (x = 0, 0.1). The targeted temperature range (2 to 300K) covers well all magnetic and structural transitions, using both thermal and cold neutron spectroscopy. This ensures accessing an extended energy and $Q$ ranges on the Stokes side, and collecting high-resolution anti-Stokes spectra. DFT-based lattice dynamical calculations in the whole Brillouin zone were performed on the stoichiometric FeTe to accompany the measurements. The INS data reveal a signature of spin-phonon coupling depicted in the observed change in the Stokes phonon spectra across the magneto-structural transitions. Moreover the high-resolution anti-Stokes spectra show a pronounced hardening of the low-energy, acoustic region of the phonon spectra, upon heating. Computationally, the exchange and correlation contributions were approximated using different functionals, with and without considering possible van der Waals (vdW) effects. The latter, presumably reflecting a lesser sensitivity of dynamical degrees of freedom than the structural ones to weak inter-layer interactions [36]. Calculated phonon spectra are found to be sensitive to magnetic degrees of freedom stemming from the magnetically active Fe sites, and to a lesser extent to vdW Te-Te interaction.

This paper is organized as follows: the experimental and computational details are provided in Section.II and Section.III, respectively. Section.IV is dedicated to the presentation and discussion of the results, and conclusions are drawn in Section.V.

## II. Experimental Details

We probed two different batches of $Fe_{1+x}Te$ with two different iron contents. The different batches were prepared using the Bridgman-Stockbarger method. Details of the synthesis technique can be found in ref. 17. The starting nominal compositions, x, of the samples were 0 and 0.1, for S1 and S2 samples, respectively. The refinement of the X-ray pattern of a third single-crystal batch with same starting composition, shows that the real interstitial iron content of the sample S1 is much higher (x = 0.08) [18]. Sample S2 belongs to the same batch as the crystal studied in ref. 15, demonstrating that the real interstitial iron content is close to that of the nominal composition (x = 0.1). The larger iron content in the second sample as compared to the first sample was confirmed by EDX. Magnetic properties of single-crystalline samples, extracted from these batches, were characterized using a Quantum Design SQUID. Figure 1 shows that the two samples exhibit different properties. The nominal composition sample, S1, exhibits a higher transition temperature with a narrow width around 70 K. This means that S1 has a low content, x, about 0.06 according to the phase diagram [13,14]. Sample S2 has a smaller transition temperature, ~ 58 K, with a broad width. According to phase diagram [13,14], the content, x, exhibits a value just above the tri-critical point, ~ 0.11. The slight difference with the reported value in ref. 16 for a crystal from the same batch lies within the uncertainty error. Consequently, the low temperature phase for both samples, S1 and S2, must be monoclinic and the magnetic ordering must be of a bi-collinear commensurate spin density wave type. The low-temperature structure was confirmed by low-temperature synchrotron single-crystal XRD on sample from batch S2 [16,17], and for another sample from a batch with the same starting composition than batch S1 [37]. Our data integrated over the elastic line also confirm this low-temperature magneto-structural assignment.

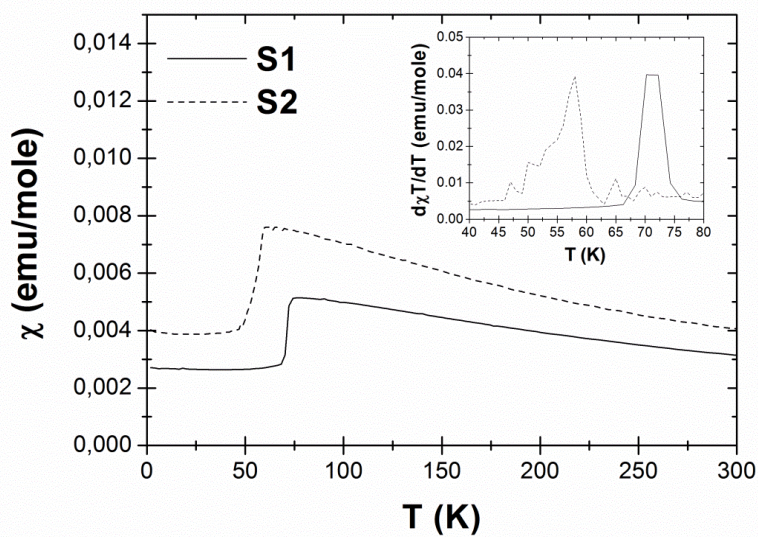

**Figure.1**: *Thermal variation of the magnetic susceptibility, χ, of single-crystal samples S1 (starting composition: FeTe) and S2 (starting composition: $Fe_{1.1}Te$). Inset: thermal variation of the temperature derivative of χT plotted in order to highlight the transition temperatures in both samples.*

The temperature-dependent INS measurements were performed at the Institut Laue-Langevin (ILL) (Grenoble, France) on ~ 2 grams of powdered samples, prepared as described above. Samples were sealed inside thin cylindrical Aluminum holder that was fixed to the cold tip of the sample stick of a standard orange cryostat. We used both the thermal, IN4C, and cold, IN6, neutron time-of-flight spectrometers. On IN4C the data were collected in the down-scattering regime (neutron energy-loss mode) using an incident neutron wavelength $\lambda_i$=1.3 Å ($E_i$=48.41 meV), offering a maximum Q-range of ~ 8.8 Å$^{-1}$, and allowing the Stokes spectrum to be measured at low temperature over a broader energy transfer range. Whereas on IN6 the excitations were probed in the up-scattering regime (neutron energy-gain mode) operating with an incident wavelength $\lambda_i$=4.14 Å ($E_i$=4.77 meV), corresponding to a maximum Q ~ 2.6 Å$^{-1}$, and offering a good resolution within the considered dynamical range for the anti-Stokes spectrum.

Standard corrections including detector efficiency calibration and background subtraction were performed. The data analysis was done using ILL procedures and software tools. The Q-averaged, one-phonon, generalized phonon density of states (GDOS) was obtained using the incoherent approximation [38-40] in the same way as in previous works dealing with phonon dynamics [41-44]. In order to prevent the contamination of phonon spectra by any eventual magnetic scattering stemming from Fe sites, only the high-Q region was considered when deriving the GDOS. Therefore, the GDOS was integrated over the Q-range 5.5 – 8.8 Å$^{-1}$ for the IN4C, low-temperature (2 - 120 K), down-scattering measurements. We extended also this approach to the IN6, high-temperature (140 – 300 K), up-scattering measurements, by considering by performing the Q-average over the momentum transfer range 1.5 - 2.6 Å$^{-1}$.

In the incoherent one-phonon approximation, the measured scattering function $S(Q,E)$, as observed in the INS experiments, is related to the phonon generalized density of states $g^n(E)$, as seen by neutrons, as follows [38-40]:

$$g^{(n)}(E) = A < \frac{e^{2W_i(Q)}}{Q^2} \frac{E}{n_T(E) + \frac{1}{2} \pm \frac{1}{2}} S(Q,E) > \qquad (1)$$

With:

$$g^{(n)}(E) = B \sum_i \left\{\frac{4\pi b_i^2}{m_i}\right\} x_i g_i(E) \qquad (2)$$

where the + or – signs correspond to energy loss or gain of the neutrons respectively and $n_T(E)$ is the Bose-Einstein distribution. $A$ and $B$ are normalization constants and $b_i$, $m_i$, $x_i$, and $g_i(E)$ are, respectively, the neutron scattering length, mass, atomic fraction, and partial density of states of the i$^{th}$ atom in the unit cell. The quantity between < > represents suitable average over all $Q$ values, within the high-Q ranges indicated above, at a given energy. $2W(Q)$ is the Debye-Waller factor. The weighting factors $\frac{4\pi b_i^2}{m_i}$ for various atoms in the units of barns/amu are [45]: Fe: 0.21 and Te: 0.034.

## III. Computational Details

Relaxed geometries and total energies were obtained using the projector-augmented wave (PAW) formalism [46,47] of the Kohn-Sham formulation of the density functional theory (KS-DFT) [48]. To estimate the effect of the exchange correlation contribution on phonon spectra in FeTe, three calculation types were done using (i) the local density approximation (LDA), (ii) the semi-local generalized gradient-corrected approximations (GGA), and (iii) the van der Waals semi-local corrected KS-DFT. The Ceperly-Alder-based parametrization by Perdew and Zunger [49] was used for the LDA calculations, whereas the GGA was approximated by the Perdew-Burke-Ernzerhof density functional scheme (PBE) [50]. We went a step further by including possible dispersive weak interactions by considering different van der Waals (vdW) based correction approximations implemented in the VASP code. This is motivated by the work of Ricci and Profeta [36] where they showed that lattice parameters, and more specifically, Fe-chalcogenide interlayer interaction is better described when considering vdW effect. Three vdW models are presently adopted; namely the Grimme's force-field correction [51] (hereafter labeled as D2), and two density functional-based vdW models, where the non-local correlation functional approximately accounts for dispersion interactions [52-54], presently in terms of the so-called vdW-optPBE and vdW-optB86b (hereafter labeled as VWPBE and VWB86, respectively).

Full geometry optimization, including cell parameters, was carried out on the experimentally refined tetragonal stoichiometric phase FeTe [18], containing two crystallographically inequivalent atoms (1 Fe and 1 Te). The space group is P4/nmm [$D_{4h}^7$] with 2 formula-units per unit cell (4 atoms). In order to determine all force constants, the supercell approach was used for lattice dynamics calculations. A supercell (3×a, 3×b, 2×c) was constructed from the relaxed geometry containing 18 formula units (72 atoms). A second partial geometry optimization (fixed lattice parameters) was performed on the supercell in order to further minimize the residual forces. The Gaussian broadening technique was adopted and all results are well converged with respect to *k*-mesh and energy cutoff for the plane wave expansion. The integration over the Brillouin zone was sampled using the Monkhorst-pack method [55]. The break conditions for the self consistent field (SCF) and for the ionic relaxation loops were set to $10^{-8}$ eV and $10^{-5}$ eV.Å$^{-1}$, respectively. This implies that Hellmann-Feynman forces following geometry optimisation were less than $10^{-5}$ eV.Å$^{-1}$. Total energies and Hellmann–Feynman forces were calculated for 8 structures resulting from individual displacements of the symmetry inequivalent atoms in the supercell, along with the inequivalent Cartesian directions ±x, ±y, and ±z. The direct method [56], as implemented in the Phonon software [57], was used to perform subsequent calculations to extract the phonon vibrational density of states, VDOS, which is transformed then to GDOS, i.e the phonon spectrum measured from INS. In contrast to VDOS, the GDOS involves a weighting of the scatterers (ions) with their scattering powers *σ/M* (*σ:* cross section and *M*: mass) [58].

# IV. Results and Discussion

Figure 2 shows the temperature dependence of the Bose factor corrected S(*Q*,E) contour plots of samples S1 and S2. Data were collected at 2, 40 and 120 K in the down-scattering regime (neutron energy-loss), using IN4C. At low temperature, no clear magnetic features can firmly be determined. This might originate from the *Q*-averaged character of the measurements performed on a powder samples over an extended *Q*-range, leading to a smearing of any magnetic signal not strong enough in nature and not decoupled from the phonon bath. Since the data were collected in the neutron energy-loss side, the spread of phonons over the whole Q-range would probably make it difficult to discriminate any magnetic signal without an eventual contamination from the phonon background. It is worth to mention the work of Stock and co-workers performed on single crystalline samples [12,13], highlighting a gapped magnetic excitation, exhibiting a softening as temperature increases. The significant magnetic scattering in their measurements was observed around Q ~ 1 Å$^{-1}$, whereas the minimum Q value in our IN4C, powder-based, INS measurements is ~ 1.4 Å$^{-1}$. We will not go further concerning magnetic excitations in the present work since presently the focus is on phonon spectra, which were derived appropriately by taking care of averaging the data over the high-Q regions of the data (cf. experimental details of the INS measurements in Section II). Upon heating, acoustic phonon dispersions are distinguishable in our measurements (Figure 2), as emanating from the Bragg spots. Phonon population follows the expected increasing trend as a function of temperature and *Q*.

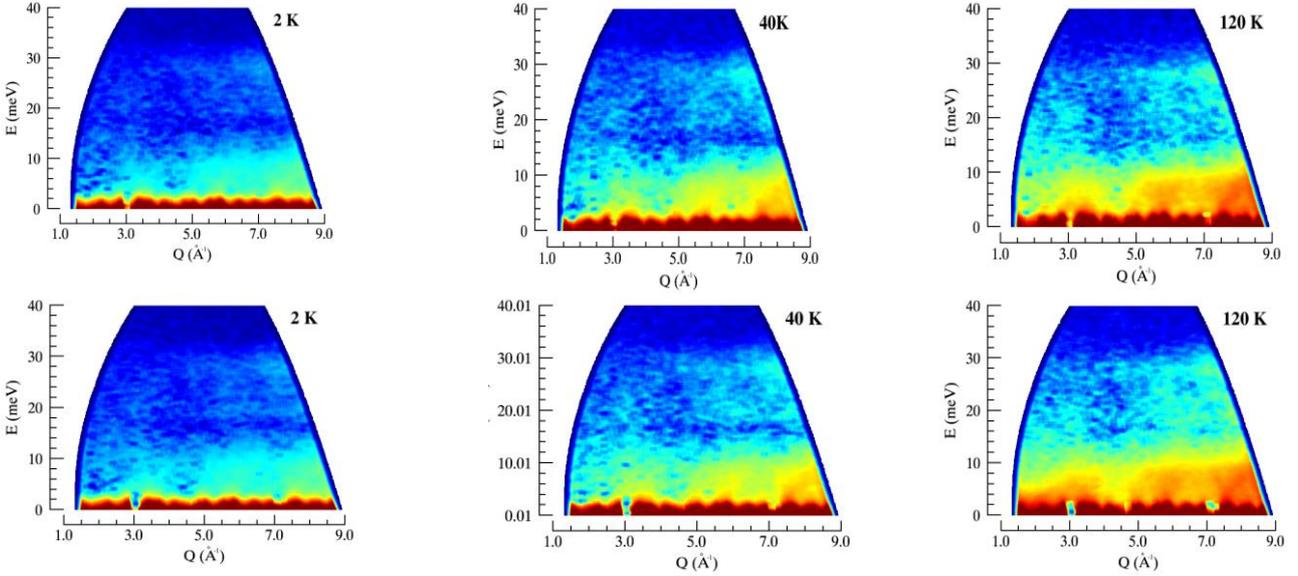

**Figure.2**: *The experimental contour plots of the temperature-dependent dynamical structure factor S(Q,E) of samples S1 (a) and S2 (b), from IN4C measurements performed in the down-scattering regime. For clarity, a logarithmic representation is used for intensities; dark red and dark blue refer to strong and weak amplitudes, respectively.*

The generalized phonon density of states (GDOS) can be derived from the measured S($Q$,E). Figure 3 shows the temperature evolution of the GDOS of samples S1 and S2. The spectra at 2 and 40 K are closely similar, or even overlapping, reflecting no effect of temperature on the phonon structure within this temperature range. Although presently we are dealing with a non-superconducting parent compound, it is worth to note that this temperature range covers the critical transition in the superconducting case. Interestingly, upon heating to 120 K, a noticeable change is induced, in both intensity and profile. A softening is observed within the energy range 15-25 meV, when temperature increases to 120 K, above the ordering temperature. The observed changes occur below and above the magnetic and structural transition temperatures, which probably points towards a spin-phonon coupling behaviour. Liu et al reported that the significant change they observed in their diffuse scattering measurements across the transition in $Fe_{1+x}Te$ was due to a large change of the lattice response to the Kanzaki forces - corresponding to a long range strain induced by the interstitial iron atoms [59]. However, a large softening of the whole phonon spectrum when heating above the transition temperature could reflect this, and this is not what we observe in the presently reported INS results.

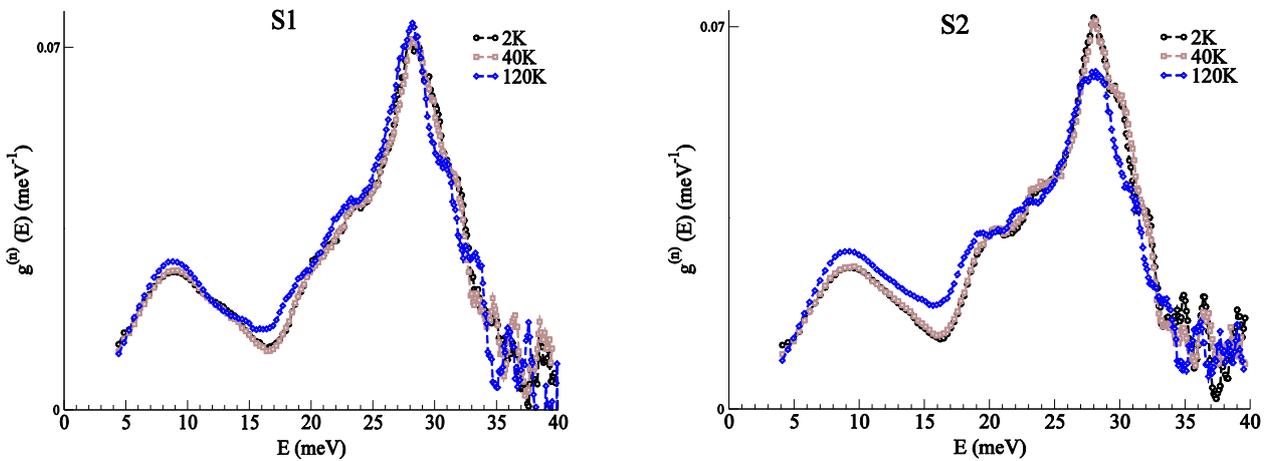

**Figure.3**: *Temperature-dependent evolution of the GDOS of samples S1 (a) and S2 (b) from IN4C measurements, performed in the down-scattering regime.*

We complement the low-temperature data by measuring high-resolution phonon spectra, in the up-scattering mode and up to 300 K, using IN6. Data were collected at 140, 220 and 300 K. Besides the high-resolution aspect, studying the temperature dependence of the anti-Stokes phonon spectra would be useful to highlight effect like anharmonicity. Figure 4 depicts the temperature evolution of phonon spectra of sample S1. We notice the non-perfect match in profile and intensity between the spectrum measured at 120 K using the thermal neutron spectrometer IN4C (Figure 3) and the spectrum probed at 140 K using the cold neutron spectrometer IN6 (Figure 4). This is because the two spectra were collected in different modes/regimes, using different neutron incident wavelengths/energies, which consequently resulted in different Q-ranges and energy resolutions (cf. experimental details of the INS measurements in Section II).

Three features are distinguishable in the low-energy part, up to 10 meV (Figure 4). First, a shoulder can be observed around 6 meV, then a peaked feature is located at 8 meV. The third remarkable observation in this energy range is a clear temperature dependence of the acoustic component, as reflected in a phonon hardening as temperature increases. The hardening amounts to about 0.6 meV, and up to 7.2 meV. In the spectral range 8-25 meV, the temperature increase leads to a slight but a detectable broadening of the features, and a decrease of their intensity. However, no shift in energy can clearly be observed within this frequency range. Between 25 and 30 meV, a clear softening with increasing temperature occurs, reflected in a frequency shift of ~ 1 meV of the peaked feature around 30 meV. The intensity of the high-energy modes around 30 meV clearly increases upon heating. The pronounced change in both position and intensity of phonon modes around 30 meV reflect, expectedly, their stretch-like nature.

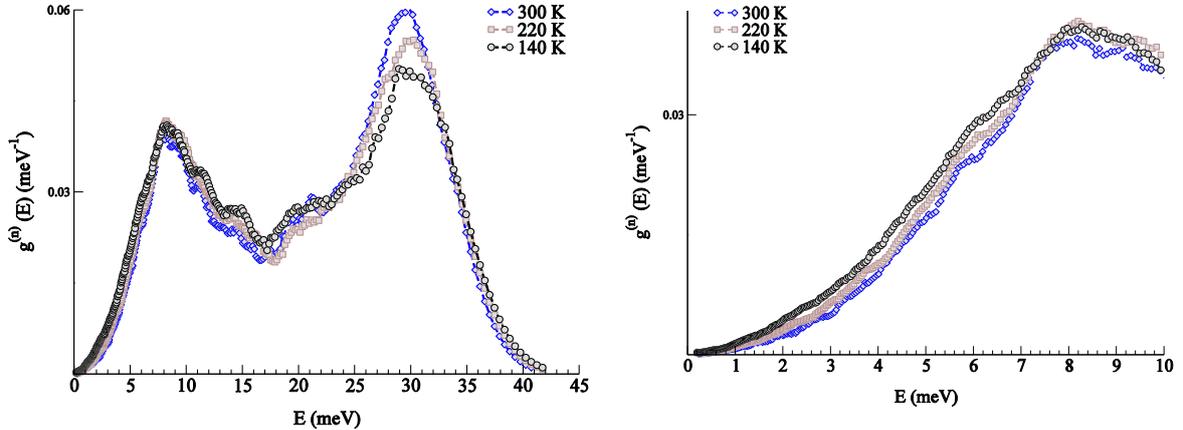

**Figure.4**: *Temperature-dependent evolution of the GDOS of sample S1 (a) from IN6 measurements, performed in the up-scattering regime. A clear temperature effect is observed within the entire energy transfer range. At low energy (b), up to 7 meV, where a weak shoulder is distinguishable, the acoustic component hardens as temperature increases. The high-energy modes, around 30 meV, exhibit a strong sensitivity upon heating, and their intensity increases with temperature.*

A comparison between measured GDOS of samples S1 and S2, at 300 K, is illustrated in Figure 5. A slight effect of the iron excess on the phonon spectra can be observed within the energy range 12 - 24 meV. However, both the spectra look closely similar in terms of intensity of the main peaked features and the frequency shifts.

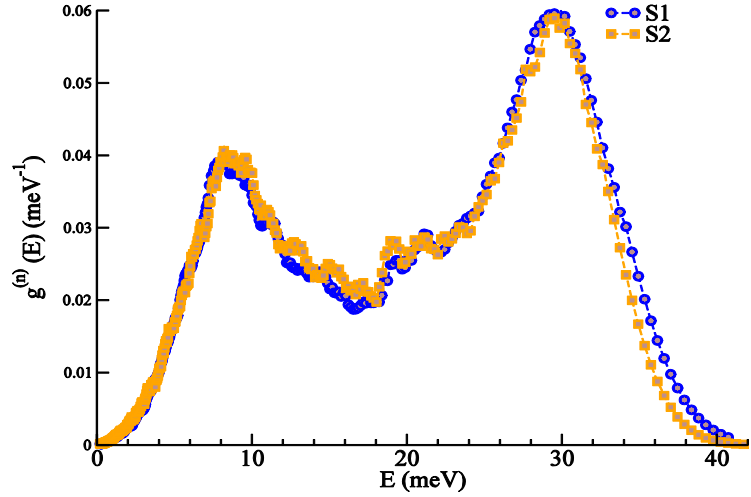

**Figure.5**: *GDOS of samples S1 and S2 at 300 K from IN6 measurements, performed in the up-scattering regime.*

The experimental results are accompanied by density functional theory based ab initio calculated phonon spectra of the tetragonal stoichiometric phase, FeTe, for a direct comparison with the high-resolution anti-Stokes phonon spectra. Calculations included different density functional methods. Spin polarization and van der Waals interaction were either considered or neglected in order to study their respective effects, and whether they act in a cooperative way or on an individual basis. In the vdW-based calculations different correction schemes as available in the VASP code were tested. In order to compare with experimental data, the calculated GDOS was determined as the sum of the partial vibrational densities of states $g_i(\omega)$ weighted by the atomic scattering cross sections and masses: $GDOS(\omega)=\Sigma_i\,(\sigma_i/M_i)g_i(\omega)$, where ($\sigma_i/M_i = 0.21$ (Fe) and $0.034$ (Te); i={Fe,Te}).

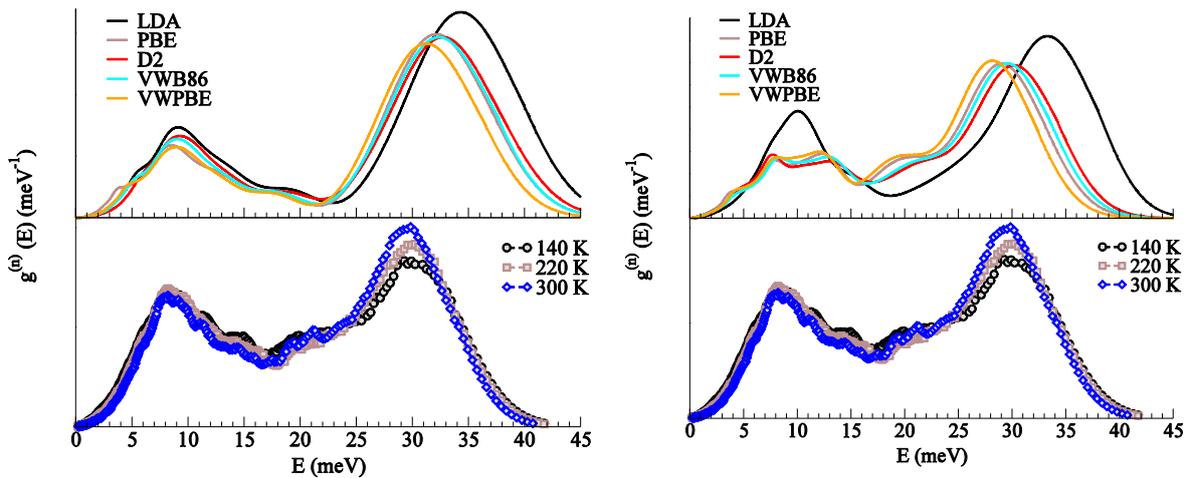

**Figure.6**: *Calculated (top panels) and measured (bottom panels) phonon spectra of sample S1. Different density functional model calculations were carried out, either by including a vdW correction (D2, VWB86, and VWPBE) or neglecting it (LDA and PBE). For each model calculation, spin-polarization was either neglected (a) or considered (b). The calculated spectra have been convolved with a Gaussian of FWHM of 10% of the energy transfer in order to describe the effect of energy resolution in the IN6 measurements.*

Figure 6 compares the ab initio determined and measured GDOS for FeTe and S1, respectively. The experimentally determined energy transfer range can be divided into three parts for the sake of the comparison with the ab initio calculations: 0 - 15 meV, 15 - 25 meV, and 25 - 35 meV.

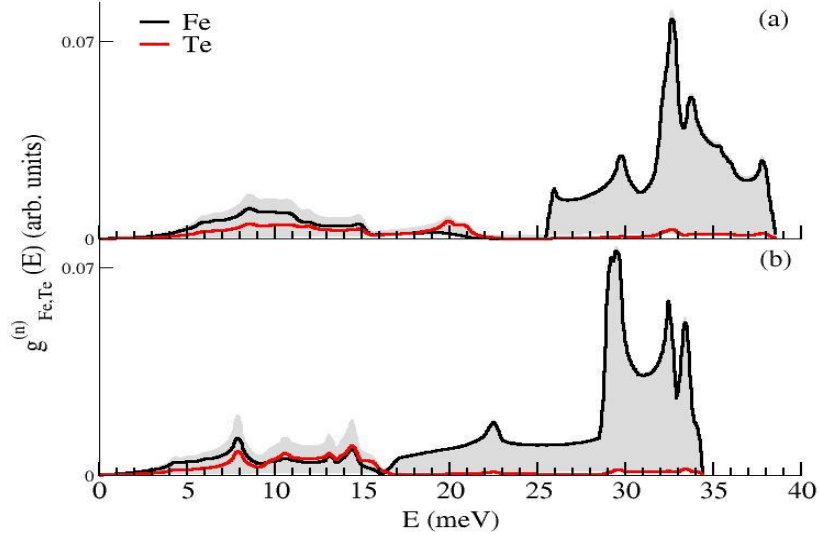

**Figure.7**: *Neutron-weighted partial phonon DOS for atoms Fe and Te in FeTe, from D2-based vdW-corrected DFT calculations. Spin degrees of freedom were either neglected (a) or included (b). The gray shaded area represents the total DOS, allowing to highlight both position and intensity contributions of the Fe and Te individual atomistic components.*

Phonon spectra from non-magnetic calculations reproduce reasonably the observed features within the first frequency range, but the agreement worsens for the second (15 - 25 meV) and the third (25 - 35 meV) range. Including vdW interactions improves only slightly that ag'reement. This demonstrates that although it was reported that dispersion forces stemming from inter-layer chalcogenide interaction was found to be important to describe structural features in Fe-based chalcogenides [36], their effect is surprisingly not strongly reflected in terms on phonon spectra, without including spin-polarization, even in the tetragonal phase. This result presents a specific dynamical aspect where the weak inter-layer interaction is overall not a dominant or a major vibrational component, without being cooperatively conjugated to the spin-polarization effect. Indeed, phonon spectra calculated within the spin-polarized framework agree better with the observations. This is not a new finding in the family of the Fe-based parent superconductors [41-44], but interestingly for the case of FeTe, as above mentioned, there seems to be a cooperative effect of considering both vdW interaction and spin-polarization. Nevertheless, to a lesser extent, the calculated phonon spectra were found to be sensitive to the vdW-based density function scheme, when restricting the comparison to the three vdW methods (D2, VWB86 and VWPBE) presently used. This confirms that the spin-phonon coupling has a strong effect on lattice dynamics of $Fe_{1+x}Te$. It is also worth to notice that our (non vdW) PBE-based phonon spectra agree with the prediction by Li and coworkers [34]. The partial atomistic phonon density of states are shown in Figure 7, where for the sake of clarity and in order to further support the above statements on the importance of spin-polarization with regards to vdW interaction, only the specific case of D2 model calculation, without (Figure 7(a)) and with spin polarization (Figure 7(b)), is depicted. In Figure 7(b), it is clear that magnetic calculations reproduce better the observation, as it can be seen, first, from the correct description of the maximum cut-off energy transfer (maximum intensity around 30 meV), which is over-estimated when spin-polarization is neglected (maximum intensity around 35 meV). Second, in Figure 7(a), the almost gaped spectral character between 15 and 25 meV, from non-magnetic calculations (without including spin-polarization effect), which experimentally should be gapless, as also confirmed by our spin-polarized calculations (Figure 7 (b)). The partial phononic contributions allow also assigning the atomistic vibrational type of the different above-mentioned spectral ranges. Indeed, considering neutron-weighted partial phonon density of states shown in Figure 7(b), motions of both Fe and Te atoms contribute to the spectra range, up to 16 meV. However, dynamics of Fe atoms dominates the mid- and high-frequency ranges, where weak interlayer interactions, involving chalcogenide, Te, are supposed to be reflected. This might explain the little effect of vdW correction on phonon dynamics in FeTe.

# V. Conclusions

In summary, we have performed temperature dependent inelastic neutron scattering (INS) measurements of the phonon spectra in two $Fe_{1+x}Te$ samples, with low interstitial iron content ($x \leq 0.11$). Both compounds, upon cooling, undergo a monoclinic to tetragonal structural transition and a paramagnetic to a bi-collinear magnetic transition. Our INS results evidence a pronounced change of the Stokes phonon spectra, signature of a spin-phonon coupling. On the other hand, the high-resolution anti-Stokes spectra reveal a pronounced hardening of the low-energy, acoustic region of the phonon spectrum, upon heating in the tetragonal phase, indicating a strong anharmonicity, and pointing towards a subtle dependence of phonons on structural evolution. The experimental results are accompanied by ab initio calculations where different density functional methods were used to account for spin-polarization and/or van der Waals interaction. Our ab-initio calculations provide a phonon density of states in good agreement with the observations. Our results suggest that including van der Waals interaction has only a slight effect on phonon dynamics. However, phonon spectra are better reproduced when spin polarization is considered, in a cooperative way with van der Waals interaction, pointing towards a pronounced spin-phonon coupling behaviour in this material. The coupling between the lattice and spin degrees of freedom highlighted in this work contributes to explain the effect of high magnetic fields or high pressures on modifying both the crystal and magnetic structures of iron tellurides [15-17], as well as it provides insights into the pronounced sensitivity to small changes in the interstitial iron contents [12,13]. Indeed, both the occurrence of irreversible magnetocrystalline domain selection observed in iron tellurides at high magnetic fields [17], and the observation of the *same* critical exponent for both structural and magnetic order parameters driving the magnetostructural transition in these compounds [13] point towards a large magnetoelastic coupling in iron tellurides which is confirmed by our present work on the lattice dynamical side. Extending INS to polarized neutron measurements offers a robust perspective offering the possibility to explore directly and unambiguously the spin-phonon coupling picture in iron telluride. In this context, the anisotropic nature of different types of low-temperature magnetic fluctuations were probed in iron telluride $Fe_{1+x}Te$ [60], confirming further the importance of magnetoelastic coupling. A next step could be the use of this same technique to draw a complete picture of both the lattice and the spin components in these materials.

# Acknowledgments

The Institut Laue-Langevin (ILL) facility, Grenoble, France, is acknowledged for providing beam time on the IN4C and IN6 spectrometers.